\documentclass[letterpaper]{article}

\usepackage{aaai2026}        
\usepackage{times}           
\usepackage{helvet}          
\usepackage{courier}         
\usepackage[hyphens]{url}    
\usepackage{graphicx}        
\usepackage{placeins}
\usepackage{natbib}          
\usepackage{caption}         
\graphicspath{{figures/}{../data/analysis/}}
\usepackage{subcaption}
\usepackage{booktabs}
\usepackage{multirow}
\usepackage{enumitem}
\usepackage{algorithm}
\usepackage{algpseudocode}
\usepackage{amsmath,amssymb}
\usepackage{csquotes}

\setlist{nosep}

\title{The Algorithmic Flattening of Sound: Computational Evidence and
Justice Implications of AI Music Homogenization}

\author{
  Zoe Slendebroek, Danaé Metaxa\\
}
\affiliations{
  University of Pennsylvania\\
  Philadelphia, United States\\
 \{zoefs, metaxa\}@seas.upenn.edu
}

\begin{document}

\maketitle

\begin{abstract}
This paper audits whether large-scale generative music systems exhibit measurable musical homogenization relative to human-produced music, and develops a justice-centered account of why this matters. 
We audit two commercially deployed systems (Suno and Lyria 3) across four genres (Afrobeats, K-pop, Dance Pop, and Heavy Metal). For each system and genre, we generate 100 tracks and compare them against human corpora of equal size, using 72 music information retrieval (MIR) features and multiple diagnostics of dispersion, redundancy, and separability. We define homogenization as reduced acoustic variation in standard computational audio features including rhythm and timing, timbre/spectral shape, and dynamics, both within genres and across genre boundaries. We also generate tracks using only a genre name as the prompt, with no additional instructions, to reveal each system's default musical tendencies. The results show two structurally distinct homogenizing tendencies. Lyria reduces within-genre acoustic diversity, while Suno collapses the acoustic distinctions between genres without compressing within-genre spread. Neither system follows user prompts faithfully, indicating that the observed patterns reflect learned priors rather than prompt constraints.
The two systems do not converge on a common acoustic profile and are more acoustically distant from each other than two random human subsamples would typically be. Nevertheless, a standard classifier distinguishes AI from human tracks near-perfectly on MIR features alone. We argue that these patterns matter not as an aesthetic curiosity but as a justice-relevant condition, shaping which musical styles become legible, valued, and economically rewarded as generated outputs increasingly circulate at scale.
\end{abstract}


\begin{links}
\end{links}

\section{Introduction}
\label{sec:introduction}

Efficiency has become a dominant organizing principle of contemporary systems. In infrastructures designed to coordinate and scale activity, improvement is increasingly equated with speed and automation. In domains such as logistics or finance, this orientation is often uncontroversial to the general public, since these systems are not intended to carry meaning in themselves. Creative work, by contrast, has historically occupied a more ambiguous position within this landscape. Its value has not been reducible to output alone, but has been bound up with authorship, interpretation, expression, and situated experience, all qualities that resist straightforward automation~\cite{manghanisavage2025}.

Generative artificial intelligence (GenAI) loosens this boundary by changing what computation can do in music and other creative tasks. Since 2023, several commercially deployed systems---including \textit{MusicLM} (Google Research), \textit{Suno} (Suno AI), and \textit{Udio} (eponymous company)---have demonstrated the ability to generate complete musical works from minimal prompts~\cite{casini2026data}. Unlike earlier computational tools, which assisted composition or automated discrete stages of production, these systems are capable of generating complete audio tracks end to end.

The increasing formal substitutability of creative labor by GenAI has prompted widespread debate across multiple domains. Legal scholarship has focused on copyright and training data~\cite{li2024copyright,ep2025genai}; industry organizations have emphasized labor displacement and economic precarity~\cite{berg}; artists and philosophers have revisited questions of authorship and authenticity~\cite{dalsgaard}. These debates are both necessary and urgent, yet they tend to center on ownership, intention, or evaluative judgments about quality, and systematic reviews of generative audio research suggest that ethical engagement with potential harms remains limited~\cite{barnett}. Comparatively little attention has been paid to how generative music systems reshape musical form when deployed at scale. This paper therefore shifts the focus from imitation or appraisal to a structural concern: \textit{do generative music systems exhibit a systematic tendency toward acoustic homogenization, and if so, what does this imply for cultural production when similarity is not merely selected for, but produced by default?}

Musical homogenization itself is not a new phenomenon. Research in the sociology of music, media studies, and cultural economics has long documented how mass distribution systems and industrial production regimes privilege musical forms that are easily legible and compatible with dominant infrastructures~\cite{adorno1941,petersonberger1975,negus2013}. Empirical studies of radio consolidation following U.S.\ telecommunications deregulation showed measurable declines in artist and stylistic diversity as playlist decisions were centralized across markets~\cite{prindle2003no}. In the streaming era, recommender systems have been shown to narrow exposure by amplifying content that performs well on engagement metrics, producing concentration even as catalogs expand~\cite{chenhuang2024}. In each case, homogenization operates structurally by rewarding familiarity and penalizing deviation, gradually aligning production with what distribution systems are most likely to surface.

What distinguishes the present moment is not the novelty of homogenizing pressures, but its relocation. Earlier mechanisms primarily operated downstream, shaping musical diversity through selection and circulation. Generative music systems extend similar optimization logics into the domain of production itself. Rather than filtering an existing repertoire, these systems generate new musical outputs by learning statistical regularities from large training corpora and sampling from learned distributions~\cite{choi2025large}. Under these conditions, homogenization is no longer only an outcome of distributional filtering and instead risks becoming a property of musical creation.

The paper proceeds in three steps. First, we situate generative music within longer histories of standardization in distribution infrastructures, production norms, and algorithmic curation. Second, we audit two commercial text-to-music systems (Suno and Lyria~3) across four genres, testing whether AI-generated outputs exhibit acoustic homogenization relative to matched human corpora. Third, building on the audit findings, we argue that this observed homogenization is not an aesthetic curiosity but a justice-relevant condition, because it can shape which musical styles become legible, valued, and economically rewarded at scale.

\noindent Accordingly, we organize the paper around the following research questions:

\begin{enumerate}
  \item \textbf{RQ1 -- Within-genre homogenization:} Do AI-generated music tracks exhibit reduced within-genre acoustic variation relative to human-produced music, and does this differ across systems?
  \item \textbf{RQ2 -- Genre boundary preservation:} Do AI systems preserve the categorical acoustic distinctions between genres, or do they collapse genre identities?
  \item \textbf{RQ3 -- System convergence:} Do AI systems converge on a shared acoustic profile in feature space, or do different systems produce structurally distinct outputs?
  \item \textbf{RQ4 -- Acoustic discriminability:} Which acoustic dimensions most reliably distinguish AI-generated from human-produced music, independent of genre?
\end{enumerate}

\noindent\textit{Contributions.}
This paper makes four contributions: it (1) develops, to our knowledge, the first AI audit approach for testing within-genre musical homogenization across commercially deployed text-to-music systems; (2) provides empirical evidence that Suno and Lyria~3 exhibit structurally distinct acoustic tendencies across four genres; (3) introduces a null-prompt baseline condition that isolates system-level priors from prompt engineering effects; and (4) develops a justice-centered analysis of how production-level homogenization can reshape cultural authority, economic value, and epistemic legitimacy in music.

\section{Related Work}
\label{sec:relatedwork}
We situate this research among other work on acoustic homogenization, before characterizing the shift introduced by generative AI, connecting this phenomenon to its implications for fairness and accountability research.

\subsection{Musical Homogenization: Concept and History}

\subsubsection{Operationalizing Musical Homogenization}
Work in cultural industries has long examined how infrastructures of production and circulation can privilege sameness over difference~\cite{adorno1941}. In this paper, we operationalize \textit{musical homogenization} in a deliberately narrow and measurable way: \textit{reduced within-genre variation} in quantifiable properties of sound, meaning that tracks become more similar to one another in standard \emph{music information retrieval (MIR)} audio features (i.e., computational descriptors of rhythm and timing, timbre, and dynamics).
This definition follows feature-based approaches to musical diversity that move beyond contested genre labels and instead measure song-to-song difference using audio descriptors~\cite{bourreau2022does}. We adopt this approach to audit whether AI-generated music exhibits lower within-genre variation compared to matched human baselines.

Empirical work on large Western music corpora demonstrates that such acoustic homogenization is historically observable. \citeauthor{serra2012}, analyzing thousands of recordings across several decades, identify declining pitch and timbral diversity alongside rising loudness and repetition. \citeauthor{mauch2015evolution} similarly demonstrate stylistic convergence using network-based analyses of harmonic and timbral features. Chart-based studies further show shifts toward slower tempos, increased prevalence of minor modes, and greater structural repetition over time~\cite{interiano2018}. Although these studies focus primarily on Western popular music, they demonstrate that acoustic homogenization can be empirically measured over time. While patterns may not generalize uniformly across musical traditions, the proposed mechanisms---including incentives embedded in production and distribution---are plausibly relevant beyond any single genre context.

\subsubsection{Historical Mechanisms of Musical Homogenization}
Prior scholarship identifies several pathways through which musical diversity is constrained by infrastructures of production and circulation~\cite{mccourtrothenbuhler1997}. Three historical mechanisms are particularly relevant: (i) industrial standardization through mass distribution, (ii) technical standardization through production norms, and (iii) algorithmic standardization through data-driven curation.

Industrial standardization constrains diversity at the level of distribution rather than creation. In the era of terrestrial radio and physical media, programming formats, genre classifications, and scheduling conventions favored songs conforming to standardized templates~\cite{mccourtrothenbuhler1997}. Following U.S.\ telecommunications deregulation in 1996, radio ownership consolidated rapidly, enabling centralized playlist control across markets and producing measurable declines in artist and stylistic diversity~\cite{dicola,prindle2003no}.

Technical standards have also reshaped musical form at the production level~\cite{vickers2010loudness}. The ``loudness war'' illustrates how competitive pressures encouraged widespread dynamic-range compression, reducing contrast between quiet and loud passages and producing a more uniform dynamic profile across recordings~\cite{vickers2010loudness}. Crucially, this shift was not driven by listener preference but by platform incentives rewarding immediate perceptual impact including radio play, playlist inclusion, and later, resistance to skip rates~\cite{meggetto}. The result was a feedback loop toward shared technical standards prioritizing competitive loudness over expressive range---an engineering choice that became industry norm and eventually acoustic signature~\cite{hinksman}.

With the migration of music consumption to digital platforms, curation increasingly operates through metric-optimized recommendation and playlist systems~\cite{prey2020curatorial,burkhart2025platformjazz}. \citeauthor{datta} demonstrate that while Spotify adoption increases consumption volume, recommendations lead to increased sales concentration, reducing diversity. \citeauthor{hosanagar} show that iTunes consumers exposed to recommendations purchase more similar titles.
Homogenization may arise not through restricting supply, but through amplified exposure to acoustically and stylistically proximal content.

Across these cases, homogenization primarily operates through \textit{selection} rather than \textit{generation}. Generative systems shift the locus of homogenization upstream, to the production rather than circulation of material.

\subsection{The Generative Turn: Homogenization at the Point of Creation}

Generative music systems introduce a novel configuration by participating directly in the production of musical material, generating audio by learning statistical regularities from large corpora and sampling from learned probability distributions~\cite{dhariwal2020jukebox}. This raises a different question than prior work on curation. What kinds of variation are reliably \textit{produced}---and which are smoothed away---when musical outputs are generated by sampling from learned distributions, rather than composed through (potentially) more heterogeneous human creative practices?

From a technical standpoint, contemporary generative models are trained to minimize loss functions that reward outputs approximating dominant patterns in training data~\cite{dhariwal2020jukebox}. Features that are frequent and stable across contexts may be reproduced more consistently than forms of variation dependent on microtiming or flexible rhythm~\cite{sturm2019}. Importantly, these systems primarily learn statistical structure in audio rather than modeling musical meaning or cultural context directly, which can favor what is computationally stable over what is musically salient in situated practice. These technical decisions may delimit which aesthetic possibilities are more likely to be learned and reproduced at scale, even in the absence of explicit normative intent on the part of the system user.

This shift has been accompanied by strong claims about democratization and creative empowerment~\cite{born,berardis}. Commercial platforms like Suno and Udio promise to make music creation available to anyone with a text prompt, regardless of training or technical skill. Reported usage figures suggest the scale of this transition. For example, Suno has publicly reported millions of users (e.g., $\sim$12M reported in 2024)~\cite{dredge2024}.

At the same time, the expressive range such systems stabilize depends on training distributions and optimization objectives. Audits of publicly available datasets used in generative music research indicate substantial geographic and cultural imbalance; one analysis estimates that approximately 94\% of training material originates from Western musical traditions, while African music accounts for roughly 0.3\%, the Middle East 0.4\%, and South Asia 0.9\%~\cite{mehta}. This imbalance is mirrored in the research ecosystem surrounding generative music. \citet{mehta} find that over 93\% of researchers in their surveyed corpus primarily focus on music from the Global North, while research contributions from Global South institutions remain substantially underrepresented. Under such conditions, statistical dominance can become difficult to disentangle from normative centrality, and low-frequency musical practices may be less likely to be learned with stability~\cite{mehta}. This mechanism provides a plausible pathway from representational imbalance to homogenization in output-level acoustic features, motivating our focus on measuring within-genre contraction and identifying which acoustic dimensions account most strongly for it.

Finally, it is worth emphasizing that the aesthetic conventions encoded in dominant training data are not culturally neutral. This type of representation problem is what \citeauthor{benjamin2019race} terms ``encoding inequality under the guise of neutrality''---i.e., the systematic privileging of certain aesthetic norms while rendering others structurally illegible. When generative models optimize for ``common elements'' in music, they reproduce patterns statistically dominant in their training corpora, which means they optimize for Western aesthetics by default~\cite{sturm2019}. With this in mind, homogenization is not simply an ``error'' in generation, but a predictable outcome of optimization under unequal representational conditions---a concern we take up in the next section.

\subsection{Why Acoustic Homogenization Matters for Fairness and Accountability}
The preceding sections establish two claims grounded in existing scholarship. First, musical homogenization has well-documented historical roots.
Second, generative systems introduce a distinct configuration by participating directly in the production of new musical material.

The remaining question is normative: why does measurable acoustic homogenization matter beyond aesthetics?

\textbf{Recognition (cultural justice).}
Cultural justice concerns the conditions under which communities can maintain, develop, and exercise control over their own cultural expressions~\cite{couldry2010}. Algorithmic homogenization becomes a cultural justice problem when platform playlists or other commercial uses elevate a narrow version of a genre as the ``representative'' one~\cite{musicai}.
Over time, this can reshape listener expectations, industry assumptions, and even pedagogical norms about what a genre is supposed to sound like~\cite{prey2020curatorial}. The problem is that these expectations may be set by platform optimization and model defaults rather than by practitioners and communities who make the music, leaving those entities with less control over how their own culture is publicly defined and remembered~\cite{drott}.

\textbf{Redistribution (economic justice).} Economic justice concerns who captures value and who bears costs under automation~\cite{rawls}. The music economy is already highly unequal, with revenue concentrated among a small set of artists and intermediaries~\cite{dicola, datta}. By enabling near-zero marginal-cost substitutes for certain kinds of musical production, generative systems can change bargaining power and licensing incentives, especially for musicians operating outside mainstream visibility~\cite{berger,drott}. Benefits may accrue disproportionately to platform owners and model developers rather than to the artists whose work constitutes training data, resembling accumulation by dispossession in which creative labor is appropriated and returned as competition~\cite{mehta}.

\textbf{Intelligibility (epistemic justice).} Epistemic justice concerns who is recognized as a knower and whose distinctions come to structure shared interpretive frameworks~\cite{felder}. In the context of generative music systems, the epistemic stakes arise from how musical knowledge is formalized, stabilized, and reproduced computationally.
As generative outputs become reference points for style and genre, model-derived regularities may be treated as authoritative, potentially subordinating practitioner knowledge about musically salient distinctions such as groove, feel, swing, regional inflection, and microtiming~\cite{danielsen,sturm2019,born}. A related risk is hermeneutical: when cultural understanding increasingly tracks what systems can represent, communities may lose vocabulary and visibility for forms of musical difference that remain perceptible in practice but are harder to formalize~\cite{dotson,mcwilliams}.

Taken together, the justice frameworks developed above suggest that algorithmic homogenization is not a niche, single-issue concern, but a multifaceted justice problem. They point toward a broader set of issues that merit sustained attention, particularly from a research community that has often treated the intersection of technology and art as a secondary ethical concern rather than a central object of analysis.

\section{Methods}
\label{sec:methods}
Below we describe our audit methodology and two experiments, followed by the analytical process used.

\subsection{Audit Scope and Systems Under Test}
\label{sec:scope}

\paragraph{Black-box, as-deployed evaluation.}

We audit two commercially deployed text-to-music systems, \textbf{Suno} (Pro tier; v5.5) and \textbf{Lyria~3}, treating both as black boxes without access to training data, architecture, or optimization objectives. All generations were collected in April--May~2026, with settings following platform defaults held constant within each condition. We applied no post-processing beyond excluding silent or corrupted outputs.

\paragraph{Genre selection.}
We audit four genres selected to provide contrastive coverage across two theoretically relevant dimensions: geographic origin (Western / non-Western) and training-data representation (high / low). These four are Afrobeats (non-Western, low), K-pop (non-Western, low), Dance Pop (Western, high), and Heavy Metal (Western, high).
We selected Afrobeats as a genre of particular justice relevance, because it contains substantial internal diversity across regional traditions, rhythmic structures, and production aesthetics~\cite{charles}, and African music is sharply underrepresented in reported training corpora ($\sim$0.3\%)~\cite{mehta}. The others were selected to balance breadth with comparability across genres: K-pop provides a large-scale non-Western comparator with higher streaming visibility. Dance Pop serves as a Western, highly represented genre. Heavy Metal occupies a very distinct acoustic space from Afrobeats, testing whether any homogenization extends across stylistically distant genres.

\paragraph{Interpretive limits.}
Findings do not estimate population prevalence across all generative music models, genres, or time periods, nor do they identify internal causes of reduced variation. Human baselines are constrained by 30-second preview availability, which may omit variation present in full tracks. Our analysis relies on standard MIR features, which may underrepresent musically salient but hard-to-formalize distinctions (e.g., groove feel, expressive microtiming). We therefore interpret results as evidence about output behavior under realistic use constraints.

\subsection{Experimental Designs}
\label{sec:designs}

We implement two complementary conditions differing in prompting strategy (Table~\ref{tab:experiments}). \textbf{Experiment~1} constructs prompts from MIR descriptors of human tracks (following guidance from the systems' own documentation) to enable \emph{one-to-one AI--human pairing} and reduce dependence on post hoc matching. \textbf{Experiment~2} uses minimal, genre-name-only prompts to isolate each system's \emph{learned prior} from prompt engineering effects, generating $n=100$ tracks per genre per system.\footnote{\textbf{Prompt examples.} Experiment~1: ``Instrumental afrobeats at 99 BPM, driving, steady groove with a tight pocket, drum-forward, loop-based and hypnotic. No vocals.'' Experiment~2: ``Instrumental Afrobeats track. No vocals, no lyrics.'' } In both conditions, AI and human excerpts are analyzed in the same standardized MIR feature space using the same homogenization diagnostics.

\begin{table}[h]
\small
\centering
\caption{Summary of experimental designs and baseline construction strategies.}
\label{tab:experiments}
\begin{tabular}{@{}p{0.16\linewidth}p{0.38\linewidth}p{0.38\linewidth}@{}}
\toprule
& \textbf{Experiment 1} & \textbf{Experiment 2} \\
\midrule
\textbf{Design} & MIR-steered prompting with one-to-one pairing & Null prompting to isolate system priors \\
\textbf{System(s)} & Suno + Lyria~3 & Suno + Lyria~3 \\
\textbf{Genre(s)} & All four genres & All four genres \\
\textbf{Prompts} & Algorithmically derived from human track MIR features & Genre-name-only (e.g., ``Instrumental Afrobeats track. No vocals.'') \\
\textbf{AI tracks} & $n=100$ per genre per system ($800$ total) & $n=100$ per genre per system ($800$ total) \\
\textbf{Human baseline} & $n=100$ tracks used to construct prompts & Same corpora as Exp.~1 \\
\textbf{Rationale} & Direct AI--human pairing; reduces matching-dependence & Isolates model prior from prompt engineering \\
\bottomrule
\end{tabular}
\end{table}

\subsubsection{Human Reference Corpora}
\label{sec:humanselection}
We curated 100 human tracks per genre by aggregating tracks from openly available Spotify playlists, filtered to release years 2000--2026. For each genre, we first collected a candidate pool of 400 tracks, then deduplicated by capping at five tracks per artist and removing duplicate titles. To represent acoustic diversity of each genre rather than reflecting playlist curation biases, we extracted MIR features from all candidate tracks and applied
k-means clustering (k=10) in the standardized feature space, then sampled proportionally across clusters to reach 100 tracks per genre. Because Spotify preview clips were unavailable at sampling time, we retrieved corresponding 30-second audio via Deezer. We further assessed potential Western curation bias by auditing artist geographic affiliations for Afrobeats using publicly available biographical information, finding substantial representation of artists embedded within African music production contexts, particularly Nigeria. For K-pop, we compared sampled artists against Billboard’s K-Pop Artist 100, which incorporates domestic Korean streaming and sales activity~\cite{billboard}; overlap with ranked Korean artists supported coverage beyond purely Western-facing popularity signals.

\subsubsection{Experiment 1: MIR-Derived Prompting (Paired Design)}
\label{sec:exp1}
For each human track, we extracted MIR features from the 30-second preview and encoded a fixed set into a natural-language prompt. To ensure prompts used vocabulary each system is designed to respond to, we followed each platform's documented prompt guidance~\cite{suno_glossary,lyria_promptguide}. For Suno, extracted feature values were translated into terms from their music glossary (e.g., BPM mapped to Italian tempo markings such as \textit{Andante} or \textit{Allegro}; onset density to texture descriptors such as sparse'' or dense''; self-similarity to loop and repetition cues). For Lyria~3, prompts were structured around the recommended elements of genre, mood, instrumentation, and tempo. All prompts requested instrumental music with no vocals. We generated one AI track per human track, giving each AI track an explicit human counterpart whose MIR profile shaped its prompt.

\subsubsection{Experiment 2: Null-Prompt Generation (System-Prior Design)}
\label{sec:exp1}
To isolate each system's learned prior from prompt engineering, we generated $n=100$ tracks per genre using only the genre name and a no-vocals instruction. The same human corpora as Experiment~1 serve as baselines. This design tests whether observed homogenization reflects the model's default learned behavior rather than an artifact of prompt construction. As a secondary output, Suno automatically assigns a title to each generated clip; we record these and analyze their diversity as a qualitative homogenization signal independent of acoustic features.

\paragraph{Audio standardization.}
All analysis used 30-second excerpts from the middle of the track for comparability across sources and to avoid intro/outro sections.

\subsection{Feature and Homogenization Diagnostics}
\label{sec:features}

We measure \textit{acoustic homogenization} as reduced within-genre acoustic variation in a standardized MIR feature space. Concretely, if AI outputs are more homogenized than human tracks, they should form a \emph{tighter} distribution (less spread) across acoustic features.

\paragraph{MIR feature set.}
We quantify acoustic similarity using 72 MIR features spanning seven interpretable domains (Table~\ref{tab:features}). Features were selected because they (1) capture acoustic dimensions that are prominent across the four genres under study---including beat timing, percussive texture, dynamic range, spectral shape, and harmonic content; (2) are widely used in MIR research with established extraction methods; and (3) enable comparison based on measurable audio properties rather than manual tagging or genre sublabels~\cite{ayodele2024afrobeat,peeters2024mir}.

\begin{table}[h]
\small
\centering
\caption{Acoustic feature domains and rationale for inclusion.}
\label{tab:features}
\begin{tabular}{@{}p{0.20\linewidth}p{0.10\linewidth}p{0.65\linewidth}@{}}
\toprule
\textbf{Domain} & \textbf{Count} & \textbf{Rationale and example features} \\
\midrule
Rhythm \& timing & 8 & All four genres are rhythmically distinctive in different ways (syncopation, groove, driving pulse). \textit{Features}: tempo, onset density, onset strength (mean, std, CV), inter-onset-interval statistics (IOI mean, std, CV). \\[4pt]

Spectral shape & 12 & Distinguishes timbral brightness, texture, and frequency distribution. \textit{Features}: centroid (mean, std, CV, var), bandwidth (mean, std), rolloff (mean, std), flatness (mean, std), spectral contrast (mean, std). \\[4pt]

MFCC timbral envelope & 30 & MFCCs capture the overall spectral envelope; delta and delta$^2$ encode the temporal dynamics of timbre — the single largest and most discriminative domain. \textit{Features}: MFCC\,0--12 (mean, std each $= 26$), first-order delta $\Delta$ (mean, std), second-order delta $\Delta^2$ (mean, std). \\[4pt]

Timbre \& texture & 4 & Captures percussive character and electronic/acoustic blend. \textit{Features}: zero-crossing rate (mean, std), harmonic ratio (HPSS), percussive ratio (HPSS). \\[4pt]

Harmonic content & 9 & Bass lines, melodic loops, and tonal identity contribute to genre character. \textit{Features}: chroma CQT (mean, std), chroma STFT (mean, std, var), chroma entropy, key clarity, Tonn\-etz (mean, std). \\[4pt]

Structure \& repetition & 3 & Loop-based production and call-response structures. \textit{Features}: self-similarity (mean, std), repetition score. \\[4pt]

Dynamics & 6 & Tests dynamic flattening (loudness war trends~\cite{vickers2010loudness}). \textit{Features}: RMS energy (mean, std, var, CV), crest factor, dynamic range (dB). \\
\bottomrule
\end{tabular}
\end{table}

All audio was processed using a single Python pipeline (\texttt{librosa}) applied identically to AI and human excerpts. Features were z-scored using means and standard deviations computed over the \emph{combined} AI+human set within each experiment to ensure cross-condition comparability. For classification analyses, scaling was performed \emph{within cross-validation folds} to prevent test-set leakage.

\paragraph{Homogenization diagnostics.}
We test for homogenization by measuring how much acoustic variation exists \emph{within} the AI tracks, and comparing it to the variation \emph{within} the human reference tracks. If AI outputs are more homogenized, the AI tracks should be more similar to one another, cluster more tightly in feature space (lower overall dispersion) and show reduced variability along individual acoustic dimensions (lower feature variance and entropy). We evaluate this pattern using five complementary diagnostics (see Table~\ref{tab:diagnostics} in the Appendix for a summary).

For global dispersion (D1), we compute each track's distance to its group center (centroid) and compare AI and human distance distributions using a one-sided Mann--Whitney U test, applied both within genre (RQ1) and across genres (RQ2). A label-shuffle permutation test (10,000 iterations) estimates how often the observed dispersion gap would arise by chance. D2 characterizes which specific acoustic features drive within-genre patterns (RQ1) by comparing feature-level variances between AI and human tracks. D3 measures whether feature distributions are more concentrated or more diffuse (RQ1). D4 tests whether AI tracks occupy a smaller or larger region of the feature space under dimensionality reduction, addressing both within-genre spread and genre boundary distinctiveness (RQ1, RQ2). D5 tests whether the two systems converge on a shared acoustic profile or remain distinct from each other and from human tracks (RQ3), and whether AI and human outputs can be reliably distinguished on MIR features alone (RQ4).

\section{Findings}
\label{sec:results}

We report findings corresponding to the four research questions. All primary analyses use Experiment~1 for both Suno and Lyria~3 across all four genres.

\subsection{Prompt Fidelity Is Low for Both Systems}
\label{sec:prompt_fidelity}

\begin{figure}[h]
    \centering
    \includegraphics[width=0.95\linewidth]{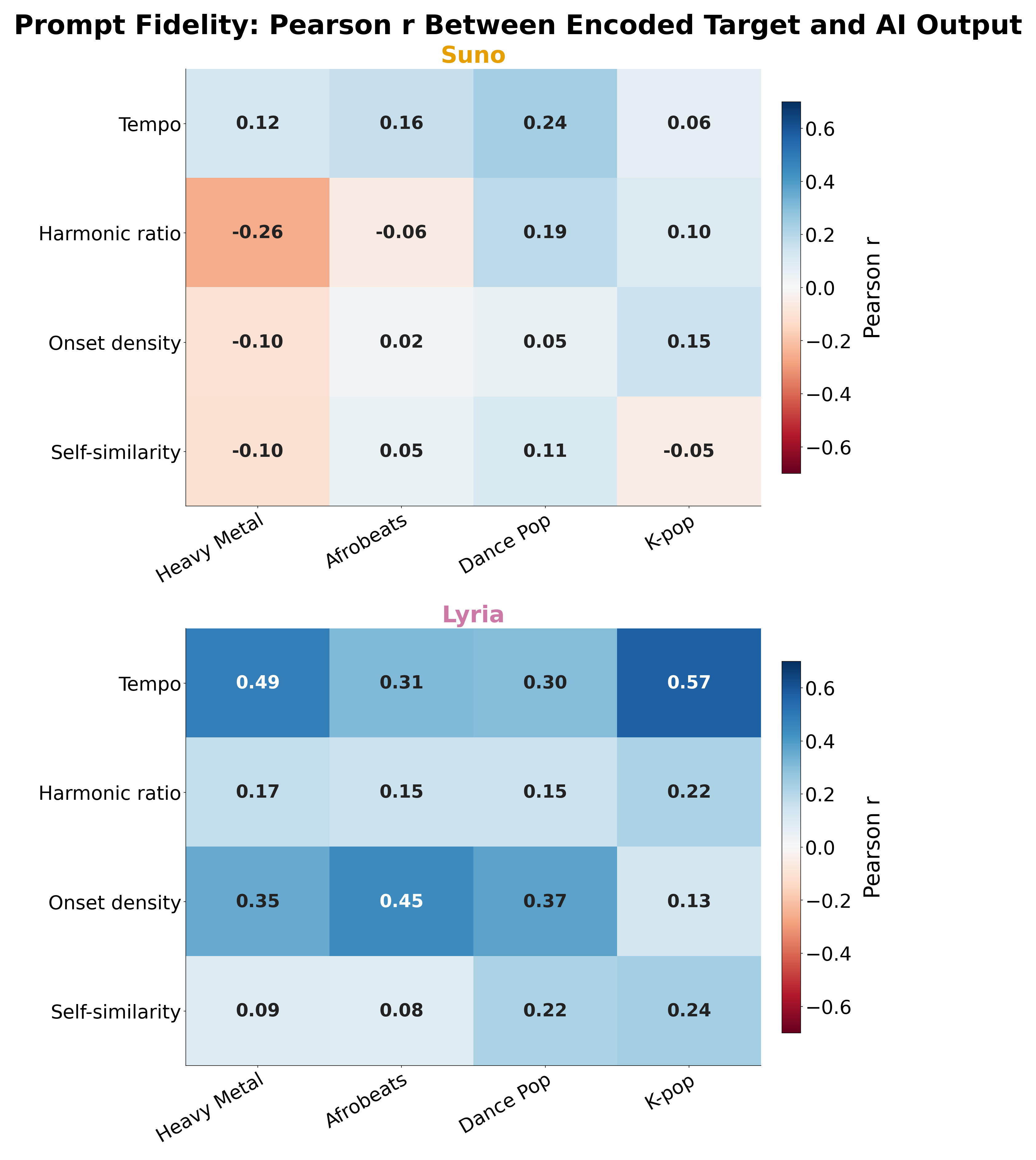}
    \caption{\textbf{Prompt fidelity: Pearson $r$ between encoded target feature value and AI output value (Experiment~1).} Suno tracks prompts only weakly ($|r|<0.26$ across all features and genres). Lyria shows moderate tempo tracking ($r=0.30$--$0.57$) but limited fidelity elsewhere.}
    \label{fig:fidelity}
\end{figure}

Before testing for homogenization, we evaluate prompt fidelity using track-level Pearson correlation ($r$) between encoded prompt features and generated outputs. Suno shows near-zero prompt fidelity across all features (max $|r|=0.26$), meaning its outputs bear almost no relationship to the acoustic properties specified in the prompt. For instance, tempo and onset density reach only $|r|=0.15$ and $|r|=0.08$ respectively, indicating the system largely ignores these instructions. Lyria tracks tempo moderately and onset density in some genres but shows limited fidelity for harmonic ratio and self-similarity (Figure~\ref{fig:fidelity}). Thus, the homogenization patterns we report are likely to reflect each system's learned priors rather than our prompt design.

\subsection{RQ1: Suno Increases Within-Genre Dispersion; Lyria Homogenizes Within Genre}
\label{sec:rq1_results}

The two systems exhibit qualitatively opposite patterns with respect to within-genre acoustic diversity.

\paragraph{Lyria homogenizes within genre.}
In three of four genres, Lyria outputs cluster more tightly in feature space than the human reference tracks. The aggregate variance ratio (AI/Human) is $0.839$, meaning AI output variance is on average $84\%$ of human variance, with $58\%$ of feature$\times$genre cells showing AI variance below human. A mixed-effects model confirms this reduction is reliable (Cohen's $d = -0.265$, $p < 0.0001$, BH-corrected). Mean pairwise distances between Lyria tracks within each genre fall below the human baseline in three of four genres (range: $0.773$--$1.119$), indicating that Lyria tracks are more similar to each other than human tracks are within the same genre (Figure~\ref{fig:diversity}; full data in Appendix Table~\ref{tab:pairwise_distances}).

\paragraph{Suno disperses within genre but collapses genre boundaries.}
Suno outputs show higher within-genre spread than human tracks (variance ratio $1.667$; $83\%$ of feature$\times$genre cells show AI $>$ human; Cohen's $d = +0.646$), with normalized pairwise distances exceeding $1.0$ in three of four genres (range: $1.004$--$1.582$). Whether this spread corresponds to meaningful genre-conditioned variation is addressed in RQ2, below.

\begin{figure}[h]
    \centering
    \includegraphics[width=1\linewidth]{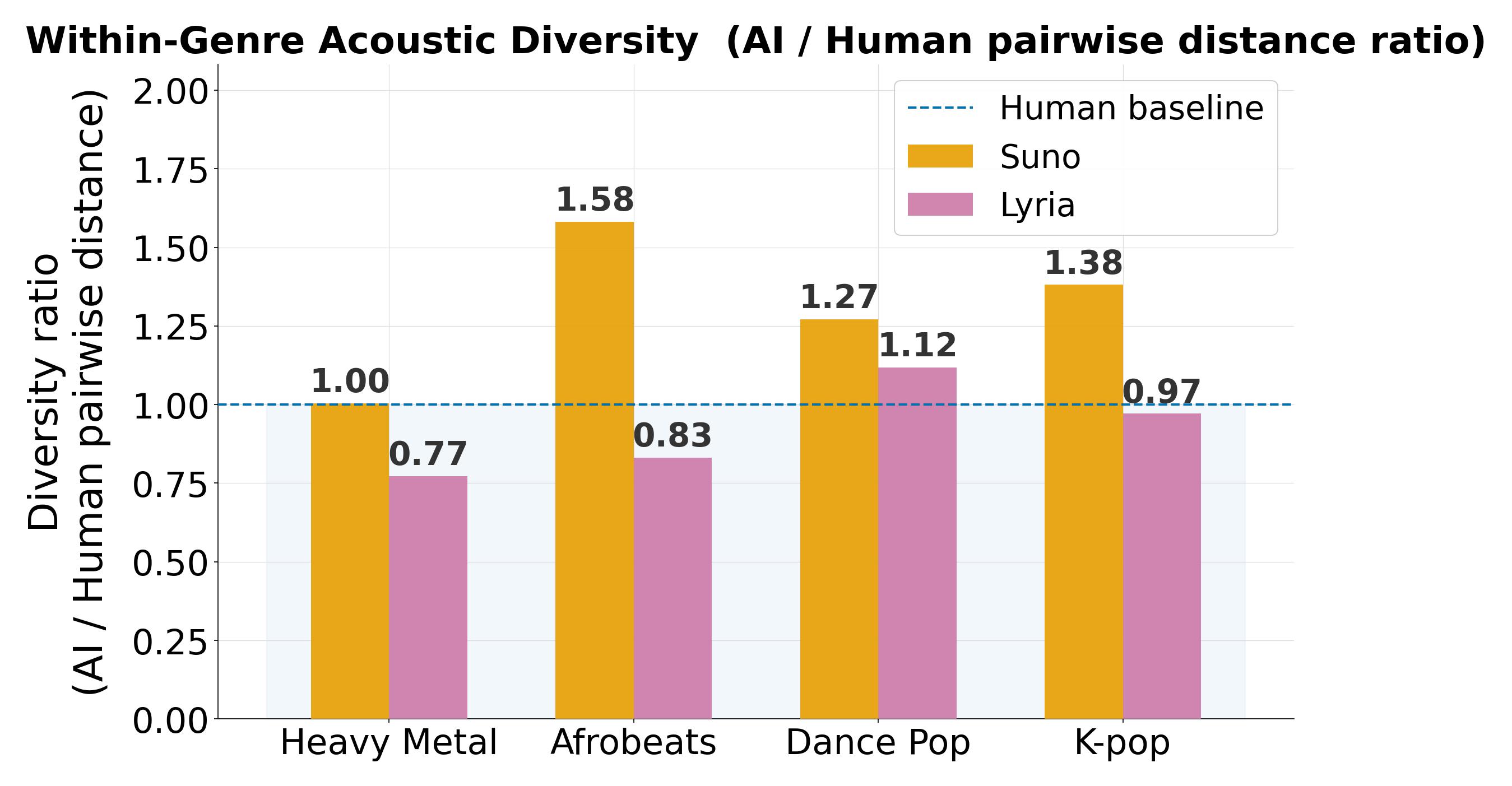}
    \caption{\textbf{Within-genre acoustic diversity by system and genre (RQ1).} Bars show the AI/Human ratio of mean pairwise distances within each genre (values $>1$ = AI more diverse than human; $<1$ = AI more homogeneous). Lyria consistently falls at or below the human baseline (blue dashed line); Suno exceeds it in three of four genres, with Afrobeats showing the largest deviation ($1.58\times$).}
    \label{fig:diversity}
\end{figure}

\subsection{RQ2: Suno Collapses Genre Boundaries; Lyria Preserves Them}
\label{sec:rq2_results}

We measure genre boundary preservation using a genre separation ratio: the distance between genre centroids divided by the average within-genre spread, computed separately for each system and for the human corpus.

\paragraph{Lyria preserves genre boundaries.}
Human tracks maintain a separation ratio of $0.662$. Lyria is nearly identical at $0.676$ ($+1\%$), indicating that Lyria's within-genre compression does not distort the overall genre structure. Despite producing outputs that are more acoustically similar to one another within each genre, Lyria reliably keeps those outputs anchored to the correct genre center.

\paragraph{Suno collapses genre boundaries.}
Suno shows a substantially reduced separation ratio of $0.429$ ($-36\%$ relative to human). Although Suno's outputs are more spread within each genre than human tracks, they do not stay acoustically anchored to their genre. Tracks bleed across genre boundaries, reducing the distinctiveness of each genre as a category (Figure~\ref{fig:genre_sep}; full data in Appendix Table~\ref{tab:genre_separation}).

\begin{figure}[h]
    \centering
    \includegraphics[width=1.1\linewidth]{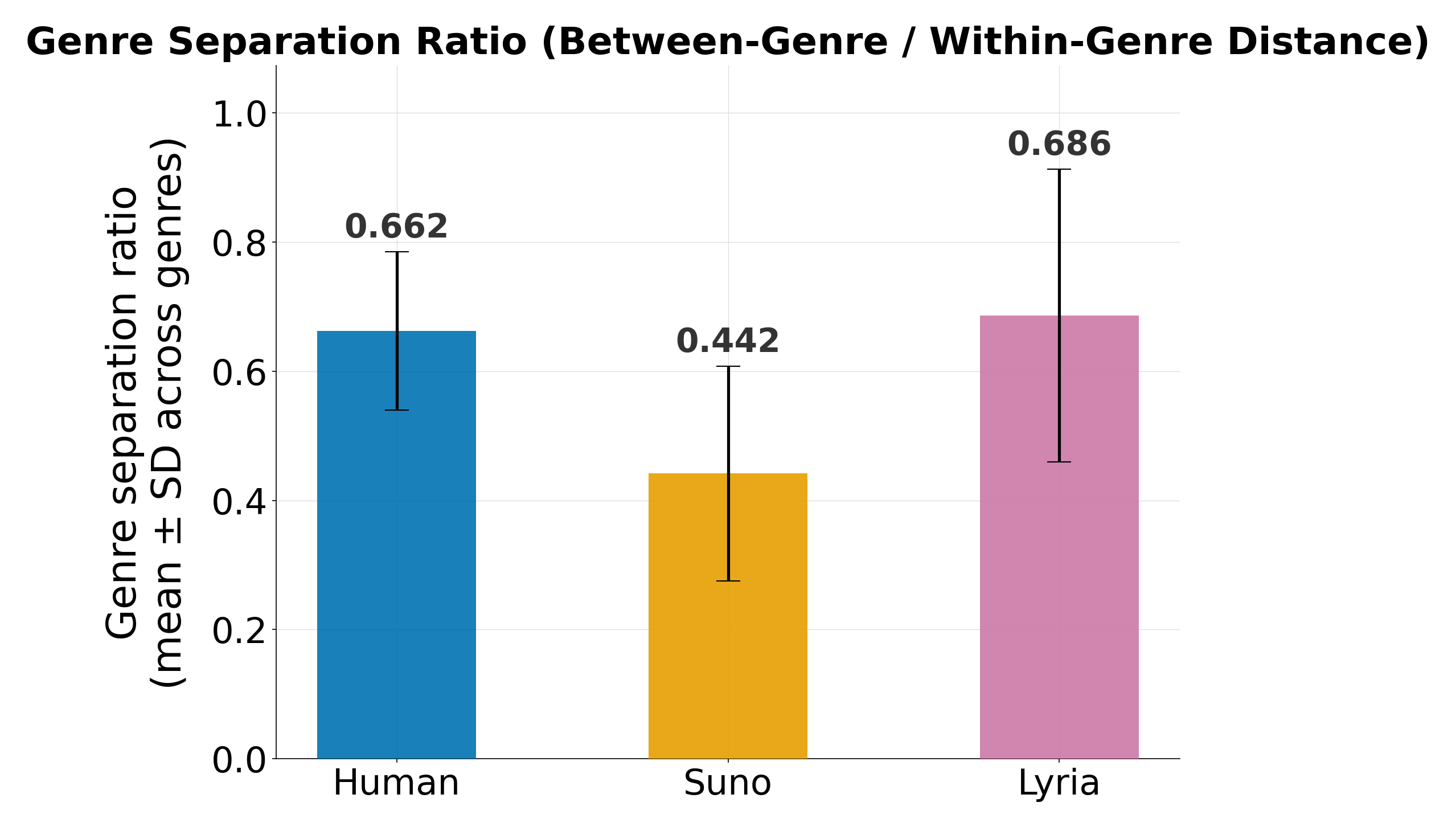}
    \caption{\textbf{Genre separation ratio by system (RQ2).} Higher values indicate more acoustically distinct genre categories. Suno's separation ratio falls $36\%$ below the human baseline, indicating that genre boundaries collapse in its output space. Lyria matches human genre separation ($+1\%$).}
    \label{fig:genre_sep}
\end{figure}

\subsection{RQ3: The Two Systems Do Not Converge on a Common Sound}
\label{sec:rq3_results}

To test whether Suno and Lyria converge toward the same acoustic output, we compute a convergence ratio: the distance between the Suno centroid and the Lyria centroid for each genre, divided by the expected distance between two random splits of the human corpus. A ratio above $1.0$ means the two systems are more different from each other than two random human subsets would be; a ratio below $1.0$ would indicate convergence.

\begin{figure}[h]
    \centering
    \includegraphics[width=1\linewidth]{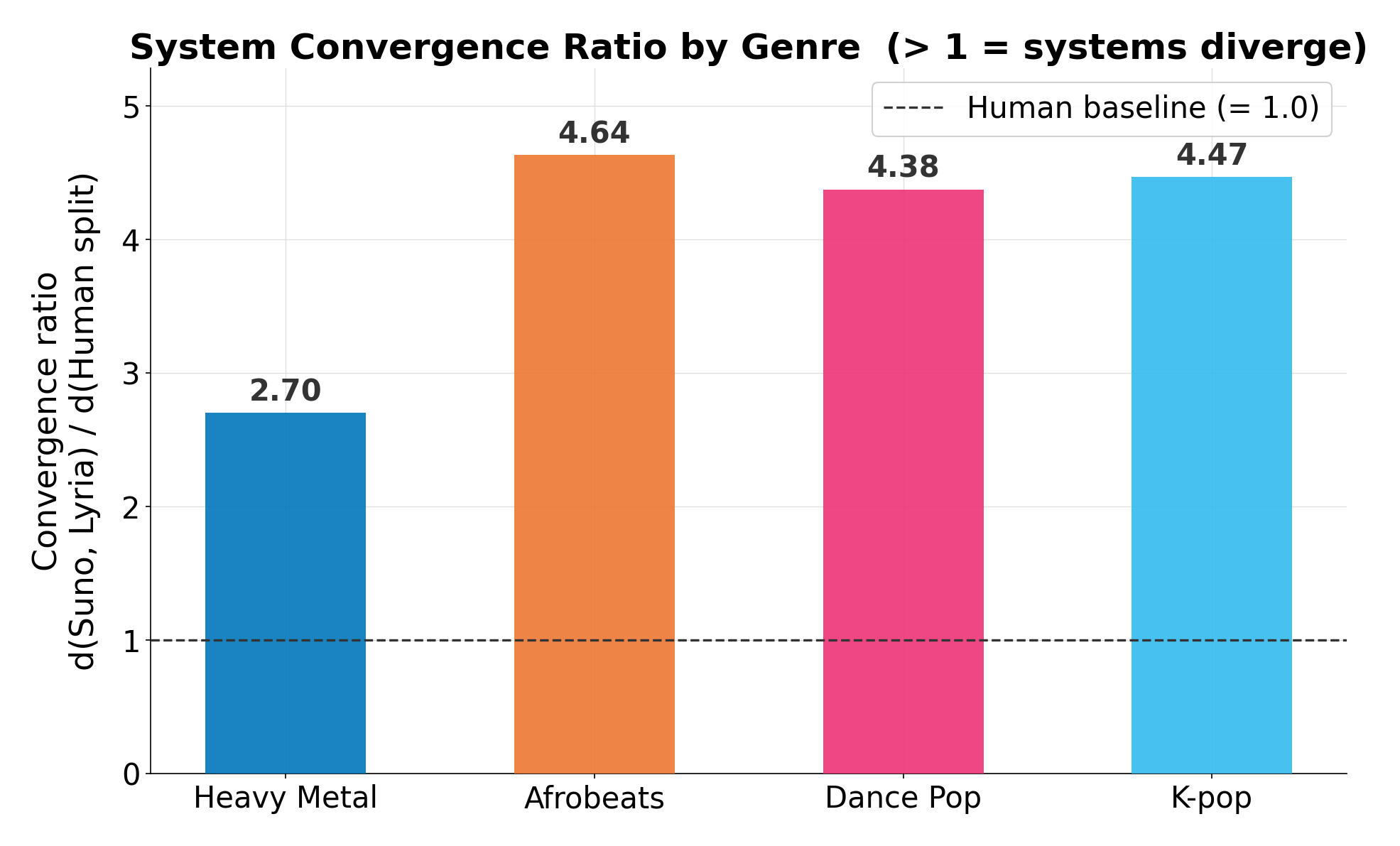}
    \caption{\textbf{System convergence ratio by genre (RQ3).} Each bar shows the distance between the Suno and Lyria centroids divided by the expected distance between two random human splits. All ratios exceed $1.0$ (dashed line), meaning the two AI systems are more acoustically different from each other than two random human subsets would be.}
    \label{fig:convergence}
\end{figure}

All ratios exceed $1.0$, ranging from $2.70$ (Heavy Metal) to $4.64$ (Afrobeats), with Dance Pop at $4.38$ and K-pop at $4.47$. The two systems are therefore more acoustically distant from each other than two random human subsets would typically be, across every genre (Figure~\ref{fig:convergence}).

\subsection{RQ4: AI and Human Outputs Are Near-Perfectly Discriminable}
\label{sec:rq4_results}

\begin{figure}[h]
    \centering
    \includegraphics[width=1\linewidth]{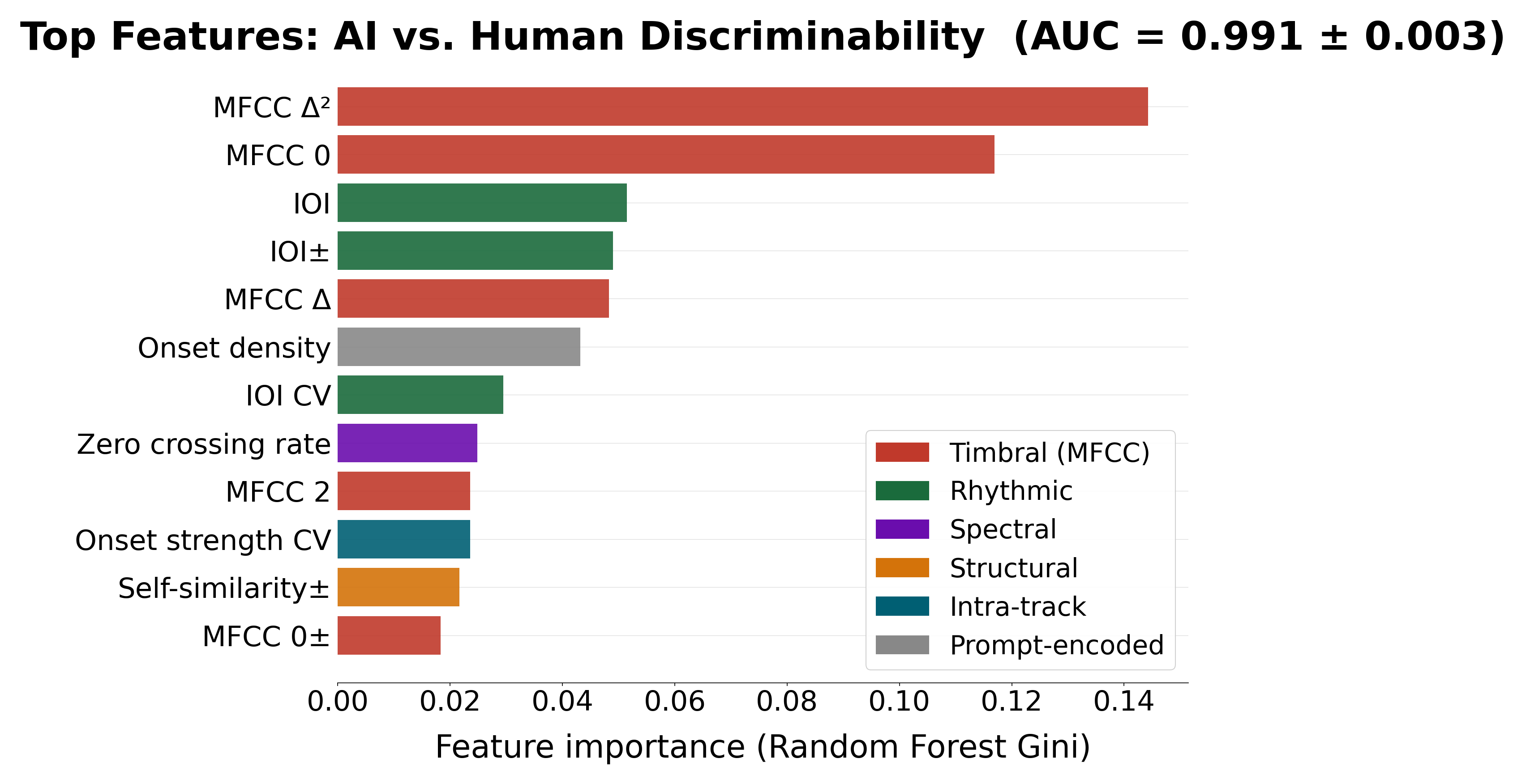}
    \caption{\textbf{Top-12 features for AI vs.\ human discriminability, Random Forest Gini importance (RQ4; AUC $=0.991\pm0.003$).} The three highest-importance features are MFCC $\Delta^2$ (timbral envelope acceleration, $0.144$), MFCC~0 (overall energy envelope, $0.117$), and IOI mean (inter-onset interval, $0.051$). Timbral and rhythmic features together account for the majority of discriminative signal. Colour indicates feature domain.}
    \label{fig:importance}
\end{figure}

We train a classifier on the 72-dimensional MIR feature set and evaluate AI-vs-human discriminability using 5-fold stratified cross-validation. The classifier separates AI from human tracks near-perfectly (mean AUC $= 0.991 \pm 0.003$; accuracy $= 0.967 \pm 0.021$). The three highest-importance features are MFCC $\Delta^2$ mean (timbral envelope acceleration, importance $= 0.144$), MFCC~0 (overall energy envelope, $0.117$), and IOI mean (average time between note onsets, $0.051$), with timbral dynamics and rhythmic regularity together accounting for the majority of discriminative signal (Figure~\ref{fig:importance}).

Notably, one potential confound is the presence of vocals in three of the four genres' human tracks. Recall that we prompted AI tracks to be instrumental to control for vocals as a potential confound, while keeping the human corpora representative of the commercial genres selected. To address this confound, we restrict the classifier to Afrobeats, the only genre where human tracks are also instrumental. Discriminability holds: AUC remained $0.980 \pm 0.011$, with IOI features gaining relative importance while vocal-sensitive MFCC features dropped from $64.5\%$ to $38.0\%$ of top-10 importance, confirming that rhythmic regularity rather than vocal content drives the separation.

\subsection{Experiment 2: Null-Prompt Baseline and Title Analysis}
\label{sec:exp2_results}

Experiment~2 uses minimal genre-name-only prompts to isolate each system's learned prior from prompting effects, testing whether the homogenization patterns observed in Experiment~1 persist when no acoustic targets are specified. Appendix Table~\ref{tab:null_comparison} reports within-genre pairwise distance ratios under null prompting alongside the MIR-steered condition.

For Suno, null-prompt and MIR-steered ratios are nearly identical (differences of $-0.302$ to $+0.094$), consistent with our earlier finding that removing acoustic targets makes no meaningful difference to what it produces. For Lyria, null-prompt tracks are more homogeneous than MIR-steered tracks in three of four genres (differences $-0.178$ to $-0.048$), in line with its partial prompt fidelity where acoustic targets expand output diversity only to the degree they are followed. Null-prompt Lyria (ratio range $0.595$--$1.071$) remains substantially more homogeneous than null-prompt Suno ($0.981$--$1.366$).

\paragraph{Title diversity as a qualitative homogenization signal.}
Suno automatically assigns a title to each generated clip, with each API call producing two clips that share a title. 
Across 100 null-prompt tracks per genre, only 13--16\% of generated titles are unique within each genre (Appendix Table~\ref{tab:titles}). A single title can account for up to 20\% of all tracks in a genre. For example, ``Palm Wine Drift'' was assigned to 20 of 100 Afrobeats tracks. Overall, diversity within individual calls is low, and the vocabulary of titles is narrow across the full 100-track set for each genre (Appendix Figure~\ref{fig:wordcloud}).

The title vocabulary also exhibits cross-genre leakage: ``Anvil Breaker'' is the top title in Heavy Metal (18 occurrences) \emph{and} appears 14 times in Dance Pop. This suggests that Suno's title generation, like its acoustic generation, collapses to a narrow prior that is not robustly conditioned on genre. Lyria~3 does not produce titles and the system treats each generation as a nameless audio object, with no representational framing attached.

\section{Discussion}
\label{sec:discussion}

The results reveal two distinct homogenization patterns. Lyria compresses tracks toward a genre center; Suno compresses genre centers toward each other. Convergence ratios of $2.70$--$4.64$ confirm that Suno and Lyria produce outputs more acoustically different from each other than two random human subsets would typically be, pointing to system-specific learned priors that override both prompting and genre conditioning. A classifier trained on the full pooled feature set distinguishes AI from human tracks near-perfectly on MIR features alone, and discriminability held in an instrumental-only test (Afrobeats tracks), confirming that vocals are not driving the separation. Rather, rhythmic timing is the primary axis of divergence when timbral cues are equated across conditions. Both patterns become consequential once generated music enters the platform infrastructure that shapes what listeners hear, how genre definitions stabilize, and what kinds of music production are economically viable. Our findings provide empirical evidence of changes in measurable acoustic properties across the evaluated datasets; however, broader implications regarding platform dynamics, cultural homogenization, and economic impacts remain interpretive and require further investigation.

\subsection{Homogenization and the Platform Feedback Loop}
\label{sec:platformFeedback}

The acoustic profiles AI systems produce by default are well suited to platform circulation. Streaming platforms rely on audio descriptors to group tracks, populate genre playlists, and optimize for engagement, a process \citet{prey2020curatorial} describes as \textit{curatorial power}, or the capacity to advance platform interests through the organization of content. Regular timing, compressed timbral variation, and stable spectral shape are precisely the properties automated classification models handle most reliably, meaning generated music is not merely entering these systems but may be especially easy for them to sort, recommend, and amplify. With Deezer reporting that $28\%$ of new uploads in 2025 were AI-generated \cite{deezer}, the scale of supply alone makes this a plausible site of platform-level influence, even if upload volume does not tell us how often such tracks are actually heard.

Genre drift is one potential implication of these patterns. No single track redefines Afrobeats or K-pop, but if AI outputs constitute a growing share of what recommendation systems surface, the reference distribution of a genre could shift toward whatever a dominant system has learned to produce. With approximately $94\%$ of generative music training data drawn from Western traditions \cite{mbzuai2025}, non-Western genres may be learned from a limited sample; notably, these are also the genres where Suno's outputs deviate furthest from the human acoustic center in our study. The cultural concern here is that, over time, such shifts could make misrepresentation feel representative. AI-generated ``artists'' in these styles have already reached millions of streams through algorithmic promotion without clear disclosure \cite{gomezsarmiento2025npr}. Research in psychology has shown that repeated exposure to a stimulus increases preference for it even when listeners are unaware of the repetition \cite{zajonc1968,bornstein1989}, so normalization may not require conscious preference; saturation alone may be sufficient to unconsciously reshape listener expectations.

A related concern is that human producers may intentionally or unintentionally begin adopting the same acoustic signatures prevalent in AI music to optimize for platform visibility and engagement; that is, homogenization may become self-reinforcing. In the future, AI systems may push genres toward the patterns they most reliably generate, while human producers may be pressured to make music that fits those patterns in order to remain visible. This could potentially result in both a technical narrowing of the feature space separating AI from human outputs, as well as a cultural narrowing of the genre itself, weakening watermarking strategies that depend on stable separation, and accelerating the genre drift described above.

\subsection{Legibility, Authority, and Cultural Justice}
\label{sec:epistemic}

The homogenization patterns documented here carry implications for what forms of musical knowledge remain legible once generation is automated, and which communities bear the cost. Groove in funk drumming, swing in jazz phrasing, and timing deviations in regional hip-hop production are musically salient within practitioner communities yet only partially captured by the spectral and rhythmic features our classifier relies on~\cite{keil1987}. The issue is therefore one of platform legibility: as streaming services organize music through playlists, categories, and recommendation systems, the parts of genre that are easiest to measure also become easiest to circulate~\cite{prey2020curatorial}. Genres then risk being represented through broad, scalable cues rather than the fine-grained distinctions practitioners use to create feel and identity. The features that survive generation at scale may be those easiest to encode, not those necessarily most musically meaningful in practice.

This asymmetry carries cultural stakes, as representational imbalances in training data determine which musical practices are learned with enough stability to be reproduced~\cite{sturm2019}. In an AI-dominated future, model-derived regularities risk being treated as authoritative accounts of a genre's sound, potentially subordinating practitioner knowledge about groove, feel, regional inflection, and microtiming~\cite{born}. When these regularities increasingly define how music is tagged, recommended, and publicly recognized, communities whose traditions are poorly represented lose interpretive authority over their own cultural forms~\cite{couldry2010,drott}. Over time, this risks a form of contributory injustice in which entire musical epistemologies are excluded from shaping how sound is computationally understood and generated~\cite{dotson}.

\subsection{Legal Stakes, Economic Displacement, and the Structural Contradiction}
\label{sec:legal}

The legal landscape surrounding AI music is already in motion. In 2024, Universal, Sony, and Warner, backed by the Recording Industry Association of America, sued Suno and Udio, alleging large-scale infringement of copyrighted sound recordings at the training stage \cite{riaa2024}. Suno has acknowledged that its training data included copyrighted recordings, while arguing that this use is protected by fair use, a question that remains unsettled in this context. The US Copyright Office has since reaffirmed that copyright requires human authorship: purely AI-generated material is not copyrightable, although works incorporating AI-generated material may qualify where sufficient human creative authorship is present \cite{uscopyright2025}. This creates a structural tension. Firms accused of building generative systems from unlicensed recordings may still operate in markets where outputs, interfaces, or human-directed uses are treated as partially ownable. Our results add a technical dimension to this tension. The outputs are not simple copies, but neither are they acoustically detached from the human musical domains on which these systems appear to rely. Their distance from human corpora is detectable and learnable, which complicates simple accounts of transformation while also showing why conventional derivative-work analysis may struggle to capture the kind of genre-level extraction and recombination at issue.

The industry response to this contradiction is not uniform. More than 200 artists---among them Billie Eilish, Nicki Minaj, Stevie Wonder, and Pearl Jam---signed an open letter in 2024 calling on AI developers to cease using their work without consent, describing the practice as an ``assault on human creativity'' that risks diluting royalty pools for working musicians \cite{ara2024}. But the position of major rights-holders is different. In a conversation we had with an executive at a major music production company in the course of conducting this work, the prevailing view he articulated was pragmatic accommodation: provided the company is compensated when its catalog is used as training data, generative tools are a potential revenue stream rather than a threat. Rights-holders with large catalogs and the legal capacity to negotiate licensing terms are positioned to benefit from that arrangement; independent and emerging artists, who own neither the catalog nor the leverage to extract compensation for its use, are not. The 2023 Hollywood writers' strike offers a useful precedent in another creative domain; the central issue is not AI use itself, but who controls its terms \cite{writers}. Music appears headed the same way, through settlements that may protect major labels more than working musicians.

The optimistic response is that listeners will still choose human-made music over AI-generated music, but much of the music economy is not governed by that preference. Background music in advertising, social media, podcasts, and ambient playlists is evaluated for functional adequacy rather than internal variation. The acoustic flatness our results document may be precisely what makes AI-generated tracks serviceable in these contexts at a cost structure no prior technology could match. 

Geographically, the asymmetry compounds the patterns identified in our audit, where genres underrepresented in training data can be replicated and sold globally while the communities that developed those styles capture little of the resulting value \cite{drott}. The music continues to circulate, yet the infrastructure no longer depends on the people who produced it.

\subsection{Creative Labor and the Democratization Claim}
\label{sec:labor}

Platforms like Suno and Lyria are widely described as democratizing music creation, enabling anyone to produce music regardless of technical skill or training. Although access effects are real, especially for hobbyists and creators in low-resource contexts where prompt-based generation can be a genuine gain, the claim carries an implicit assumption: that the output is what matters and that the process of arriving at it is incidental. Access to a finished audio file and access to musical craft are two different things, with a substantial body of work in music psychology and education treating musical production as a developmental, social, and expressive process whose value is not transferred to someone who simply receives a generated artifact~\cite{hallam}.

The structural concern is what happens at scale to the conditions under which musical knowledge is developed and transmitted. If production speed increases by an order of magnitude, it could alter demand for some forms of production work and affect existing pathways for for developing expertise. The optimistic counterpoint, that human music becomes more valuable as AI output homogenizes, is structurally plausible but sits against current incentive structures. The commercial pressure is to reduce cost by producing music that is functionally adequate rather than distinctive, and nothing in the patterns our audit documents suggests those pressures are working in the other direction.

\subsection{Limitations \& Future Work}
\label{sec:limitations}

There are several limitations to consider in the context of this work. Our study audits two commercially deployed systems across four genres under specific prompting regimes, but does not identify causal mechanisms behind the observed patterns, a common constraint of black-box audits \cite{metaxa}. The results characterize output behavior under realistic use conditions rather than the upper bound of what more sophisticated prompting might achieve; prompt fidelity is limited for both systems, and expert prompting could partially alter the patterns observed. These findings nonetheless reflect how most users encounter these systems, since both platforms are designed and marketed for use without musical training or expertise.

In our experiments, human reference corpora are constrained to 30-second previews, which may omit variation present in full tracks and could understate within-genre human diversity. However, a significant confound exists in the vocal/instrumental dimension, as AI-generated tracks are purely instrumental while human corpus tracks may contain vocals, which could inflate AI--human separation on timbre-linked features. We address this directly by restricting the classifier to Afrobeats, the only genre with a fully instrumental human corpus, but further work is needed to confirm this on a wider set of genres. 

Finally, the MIR feature set upon which we base our analyses, though widely used, reflects dominant Western production norms and may underrepresent musically salient distinctions in non-Western traditions~\cite{peeters2024mir}---a limitation that is particularly relevant for Afrobeats and K-pop, the genres where cultural stakes are highest and where our audit is most justice-relevant. Future work should complement computational audits with listener studies examining whether these measured acoustic differences correspond to perceived changes in genre identity and  authenticity.

\section{Conclusion}
\label{sec:conclusion}

Generative AI has received sustained attention across industries for questions of creativity and authorship. Expanding on these ideas, this paper asks: how does AI generation in music impact the music being created and heard? We audit whether AI music generation leads to measurable homogenization, especially in genres underrepresented in training corpora, and develop a justice-centered account of why this matters once AI-generated music circulates at scale.
We conducted an audit of two commercially deployed text-to-music systems (Suno and Lyria~3) across four genre contexts (Afrobeats, K-pop, Dance Pop, and Heavy Metal), using two complementary experiments: one with prompts derived from acoustic descriptors of matched human tracks, and one with minimal genre-name-only prompts. The two systems exhibit structurally distinct patterns: Lyria reduces within-genre acoustic diversity, while Suno collapses the acoustic distinctions between genres without compressing within-genre spread. Despite diverging from each other, both systems produce outputs that are readily distinguishable from human-created music on acoustic features alone.

We argue that these patterns are not just an aesthetic curiosity. If AI-generated music is more internally uniform, it may be especially compatible with the same platform infrastructures that sort tracks into genres and playlists and reward what is predictable and easy to categorize. This could also shift public expectations toward the platform-legible version of a genre, smoothing away the internal variation that human musicians use to create feel and identity. The concerns are especially relevant for styles that are culturally underrepresented in training data, where a system may reproduce a genre in a simplified and standardized form that is easier to scale. We call, then, not for the wholesale rejection of generated music, but for a deliberate reckoning with what forms of variation, diversity, and human involvement we value in the music we consume.

\appendix

\bibliography{references}

@article{adorno1941,
  title={On Popular Music},
  author={Adorno, Theodor W.},
  journal={Studies in Philosophy and Social Science},
  volume={9},
  number={1},
  pages={17--48},
  year={1941}
}

@article{ayodele2024afrobeat,
  title={The influence of african rhythms on modern music: A case study of afrobeat in Nigeria},
  author={Ayodele, T},
  journal={Art and Society},
  volume={3},
  number={1},
  pages={45--52},
  year={2024}
}

@inproceedings{barnett,
  title={The ethical implications of generative audio models: A systematic literature review},
  author={Barnett, Julia},
  booktitle={Proceedings of the 2023 AAAI/ACM Conference on AI, Ethics, and Society},
  pages={146--161},
  year={2023}
}

@article{benjamin2019race,
  title={Race After Technology: Abolitionist Tools for the New Jim Code},
  author={Benjamin, Ruha},
  journal={Polity},
  volume={1499},
  year = 2019
}

@article{berardis,
  title={Towards responsible AI Music: An investigation of trustworthy features for creative systems},
  author={de Berardinis, Jacopo and Porcaro, Lorenzo and Merono-Penuela, Albert and Cangelosi, Angelo and Buckley, Tess},
  journal={arXiv preprint arXiv:2503.18814},
  year={2025}
}

@article{berg,
  title={Generative AI and the media and culture industry},
  author={Berg, Jenine and Gmyrek, P and Licata, M and Gwenyaya, T and Scaiano, MSF},
  journal={ILO Research Brief},
  number={1},
  year={2025}
}

@misc{berger,
  title={AI’s impact on music in 2025: Licensing, creativity, and industry survival},
  author={Berger, V},
  year={2024},
  publisher={Forbes. https://www. forbes. com/sites/virginieberger/2024/12/30/ais-impact~…}
}

@online{billboard,
  author = {{Billboard Korea}},
  title = {K-Pop Artist 100 2025},
  year = {2025},
  url = {https://www.billboard.com/lists/k-pop-artist-100-2025/},
  urldate = {2026-08-05}
}

@article{born,
  title={Artificial intelligence, music recommendation, and the curation of culture},
  author={Born, Georgina and Morris, Jeremy and Diaz, Fernando and Anderson, Ashton},
  year={2021}
}

@article{bourreau2022does,
  title={Does digitization lead to the homogenization of cultural content?},
  author={Bourreau, Marc and Moreau, Fran{\c{c}}ois and Wikstr{\"o}m, Patrik},
  journal={Economic Inquiry},
  volume={60},
  number={1},
  pages={427--453},
  year={2022},
  publisher={Wiley Online Library}
}

@book{burkhart2025platformjazz,
  title={Platform Jazz: Algorithmic Music Culture on TikTok},
  author={Burkhart, Benjamin},
  year={2025},
  publisher={transcript Verlag}
}

@misc{suno_glossary,
  author       = {{Suno}},
  title        = {Music Glossary for {Suno}},
  year         = {2025},
  howpublished = {\url{https://help.suno.com/en/articles/9010177}},
  note         = {Accessed: 2026-04-01}
}

@misc{lyria_promptguide,
  author       = {{Google}},
  title        = {Music Generation Prompting Guide for {Lyria}},
  year         = {2025},
  howpublished = {\url{https://cloud.google.com/vertex-ai/generative-ai/docs/music/music-gen-prompt-guide}},
  note         = {Accessed: 2026-04-01}
}

@article{casini2026data,
  title={Data-Driven Analysis of Text-Conditioning in AI-Generated Music: A Case Study with Suno and Udio},
  author={Casini, Luca and Vila, Laura Cros and Dalmazzo, David and Kaila, Anna-Kaisa and Sturm, Bob LT},
  journal={Transactions of the International Society for Music Information Retrieval},
  volume={9},
  number={1},
  year={2026}
}

@article{charles,
  title={Black on Black sounds: Music, Migration, and the ‘NU-K Blak’identity formation in early 21st century Britain},
  author={Charles, Monique},
  journal={IASPM Journal},
  volume={15},
  number={2},
  pages={21--43},
  year={2025}
}

@article{chenhuang2024,
  title={Effective content recommendation in new media: Leveraging algorithmic approaches},
  author={Chen, Yi and Huang, Jueru},
  journal={IEEE Access},
  volume={12},
  pages={90561--90570},
  year={2024},
  publisher={IEEE}
}

@article{choi2025large,
  title={Large-scale training data attribution for music generative models via unlearning},
  author={Choi, Woosung and Koo, Junghyun and Cheuk, Kin Wai and Serr{\`a}, Joan and Mart{\'\i}nez-Ram{\'\i}rez, Marco A and Ikemiya, Yukara and Murata, Naoki and Takida, Yuhta and Liao, Wei-Hsiang and Mitsufuji, Yuki},
  journal={arXiv preprint arXiv:2506.18312},
  year={2025}
}

@article{couldry2010,
  title={Why voice matters: Culture and politics after neoliberalism},
  author={Couldry, Nick},
  year={2010},
  publisher={Sage}
}

@inproceedings{dalsgaard,
  title={GenAI and the crisis of creative labor: Automation, augmentation, and the artist’s role},
  author={Dalsgaard, Peter},
  booktitle={Adjunct Proceedings of the Sixth Decennial Aarhus Conference: Computing X Crisis},
  pages={1--5},
  year={2025}
}

@article{danielsen,
  title={There’s more to timing than time: Investigating musical microrhythm across disciplines and cultures},
  author={Danielsen, Anne and Br{\o}vig, Ragnhild and B{\o}hler, Kjetil Klette and C{\^a}mara, Guilherme Schmidt and Haugen, Mari Romarheim and Jacobsen, Eirik and Johansson, Mats S and Lartillot, Olivier and Nymoen, Kristian and Oddekalv, Kjell Andreas and others},
  journal={Music Perception: An Interdisciplinary Journal},
  volume={41},
  number={3},
  pages={176--198},
  year={2024},
  publisher={University of California Press}
}

@article{datta,
  title={Changing their tune: How consumers’ adoption of online streaming affects music consumption and discovery},
  author={Datta, Hannes and Knox, George and Bronnenberg, Bart J},
  journal={Marketing Science},
  volume={37},
  number={1},
  pages={5--21},
  year={2018},
  publisher={INFORMS}
}

@article{dhariwal2020jukebox,
  title={Jukebox: A generative model for music},
  author={Dhariwal, Prafulla and Jun, Heewoo and Payne, Christine and Kim, Jong Wook and Radford, Alec and Sutskever, Ilya},
  journal={arXiv preprint arXiv:2005.00341},
  year={2020}
}

@article{dicola,
  title={False Premises, False Promises},
  author={DiCola, Peter},
  journal={Future of Music Coalition, December},
  year={2006}
}

@article{dotson,
  title={Conceptualizing epistemic oppression},
  author={Dotson, Kristie},
  journal={Social epistemology},
  volume={28},
  number={2},
  pages={115--138},
  year={2014},
  publisher={Taylor \& Francis}
}

@misc{dredge2024,
  author       = {Dredge, Stuart},
  title        = {Suno Releases Its First Mobile App after Attracting 12m Users},
  year         = {2024},
  howpublished = {\url{https://musically.com/2024/07/03/suno-releases-its-first-mobile-app-after-attracting-12m-users/}},
  note         = {Accessed: 2026-04-01}
}

@article{drott,
  title={Copyright, compensation, and commons in the music AI industry},
  author={Drott, Eric},
  journal={Creative Industries Journal},
  volume={14},
  number={2},
  pages={190--207},
  year={2021},
  publisher={Taylor \& Francis}
}

@article{zajonc1968,
  title={Attitudinal effects of mere exposure.},
  author={Zajonc, Robert B},
  journal={Journal of personality and social psychology},
  volume={9},
  number={2p2},
  pages={1},
  year={1968},
  publisher={American Psychological Association}
}

@misc{mbzuai2025,
  author        = {Mehta, Atharva and others},
  title         = {Music for All: Representational Bias and Cross-Cultural Adaptability of Music Generation Models},
  year          = {2025},
  eprint        = {2502.07328},
  archivePrefix = {arXiv},
  primaryClass  = {cs.SD}
}

@article{ep2025genai,
  title={Generative AI and copyright: Training, creation, regulation. Policy Department for Justice},
  author={Lucchi, Nicola},
  journal={Civil Liberties and Institutional Affairs, Directorate-General for Citizens’ Rights, Justice and Institutional Affairs, PE},
  volume={774},
  year={2025}
}

@article{felder,
  title={Bridging the gap between knowledge and justice: Epistemic challenges in participatory disability research},
  author={Felder, Franziska},
  journal={Theory and Research in Education},
  volume={23},
  number={3},
  pages={268--283},
  year={2025},
  publisher={Sage Publications Sage UK: London, England}
}

@misc{gomezsarmiento2025npr,
  author       = {Gomez Sarmiento, Ignacio},
  title        = {{AI}-Generated Music Is Here to Stay. Will Streaming Services Like {Spotify} Label It?},
  year         = {2025},
  howpublished = {\url{https://www.npr.org/2025/08/08/nx-s1-5492314/ai-music-streaming-services-spotify}},
  note         = {Accessed: 2026-04-01}
}

@misc{riaa2024,
  author       = {{RIAA}},
  title        = {Record Companies Bring Landmark Cases for Responsible {AI} Against {Suno} and {Udio} in {Boston} and {New York} Federal Courts, Respectively},
  year         = {2024},
  howpublished = {\url{https://www.riaa.com/record-companies-bring-landmark-cases-for-responsible-ai-againstsuno-and-udio-in-boston-and-new-york-federal-courts-respectively/}},
  note         = {Accessed: 2026-05-15}
}

@article{bornstein1989,
  title={Exposure and affect: Overview and meta-analysis of research, 1968--1987.},
  author={Bornstein, Robert F},
  journal={Psychological bulletin},
  volume={106},
  number={2},
  pages={265},
  year={1989},
  publisher={American Psychological Association}
}

@phdthesis{hinksman,
  title={Creative mastering: A new culture of audio post-production},
  author={Hinksman, Alexander},
  year={2022},
  school={Birmingham City University}
}

@article{hosanagar,
  title={Will the global village fracture into tribes? Recommender systems and their effects on consumer fragmentation},
  author={Hosanagar, Kartik and Fleder, Daniel and Lee, Dokyun and Buja, Andreas},
  journal={Management Science},
  volume={60},
  number={4},
  pages={805--823},
  year={2014},
  publisher={INFORMS}
}

@article{interiano2018,
  title={Musical trends and predictability of success in contemporary songs in and out of the top charts},
  author={Interiano, Myra and Kazemi, Kamyar and Wang, Lijia and Yang, Jienian and Yu, Zhaoxia and Komarova, Natalia L},
  journal={Royal Society open science},
  volume={5},
  number={5},
  year={2018},
  publisher={The Royal Society}
}

@article{keil1987,
  title={Participatory discrepancies and the power of music},
  author={Keil, Charles},
  journal={Cultural anthropology},
  volume={2},
  number={3},
  pages={275--283},
  year={1987},
  publisher={JSTOR}
}

@article{li2024copyright,
  title={Copyright protection during the training stage of generative AI: Industry-oriented US law, rights-oriented EU law, and fair remuneration rights for generative AI training under the UN's international governance regime for AI},
  author={Li, Kaigeng and Wu, Hong and Dong, Yupeng},
  journal={Computer Law \& Security Review},
  volume={55},
  pages={106056},
  year={2024},
  publisher={Elsevier}
}

@article{manghanisavage2025,
  title={Rigorous creativity: AI, art and electronic life},
  author={Manghani, Sunil and Savage, Tom},
  journal={Journal of Visual Art Practice},
  volume={24},
  number={4},
  pages={427--449},
  year={2025},
  publisher={Taylor \& Francis}
}

@article{mauch2015evolution,
  title={The evolution of popular music: USA 1960--2010},
  author={Mauch, Matthias and MacCallum, Robert M and Levy, Mark and Leroi, Armand M},
  journal={Royal Society open science},
  volume={2},
  number={5},
  year={2015},
  publisher={The Royal Society}
}

@article{mccourtrothenbuhler1997,
  title={SoundScan and the consolidation of control in the popular music industry},
  author={McCourt, Tom and Rothenbuhler, Eric},
  journal={Media, Culture \& Society},
  volume={19},
  number={2},
  pages={201--218},
  year={1997},
  publisher={Sage Publications}
}

@article{mcwilliams,
  title={Testimonial Injustice and the Nature of Epistemic Injustice},
  author={McWilliams, Emily Colleen},
  journal={A Companion to Epistemology},
  volume={1},
  pages={557--568},
  year={2025},
  publisher={Wiley Online Library}
}

@inproceedings{meggetto,
  title={Why people skip music? On predicting music skips using deep reinforcement learning},
  author={Meggetto, Francesco and Revie, Crawford and Levine, John and Moshfeghi, Yashar},
  booktitle={Proceedings of the 2023 Conference on Human Information Interaction and Retrieval},
  pages={95--106},
  year={2023}
}

@article{mehta,
  title={Missing melodies: Ai music generation and its" nearly" complete omission of the global south},
  author={Mehta, Atharva and Chauhan, Shivam and Choudhury, Monojit},
  journal={arXiv preprint arXiv:2412.04100},
  year={2024}
}

@article{metaxa,
  title={Auditing algorithms: Understanding algorithmic systems from the outside in},
  author={Metaxa, Dana{\"e} and Park, Joon Sung and Robertson, Ronald E and Karahalios, Karrie and Wilson, Christo and Hancock, Jeff and Sandvig, Christian},
  journal={Foundations and Trends{\textregistered} in Human--Computer Interaction},
  volume={14},
  number={4},
  pages={272--344},
  year={2021},
  publisher={Emerald Publishing Limited}
}

@misc{deezer,
  author       = {{Music Ally}},
  title        = {Deezer Says 28\% of Its New Music Uploads Are Now {AI}-Generated},
  year         = {2025},
  howpublished = {\url{https://musically.com/2025/09/11/deezer-says-28-of-its-new-music-uploads-are-now-ai-generated/}},
  note         = {Accessed: 2026-04-01}
}

@article{musicai,
  title={Music and Artificial Intelligence: Artistic Trends},
  author={Pons, Jordi and Zukowski, Zack and Parker, Julian D and Carr, CJ and Taylor, Josiah and Evans, Zach},
  journal={arXiv preprint arXiv:2508.11694},
  year={2025}
}

@book{negus2013,
  title={Music genres and corporate cultures},
  author={Negus, Keith},
  year={2013},
  publisher={Routledge}
}

@misc{peeters2024mir,
  author        = {Peeters, Geoffroy and Rafii, Zafar and Fuentes, Magdalena and Duan, Zhiyao and Benetos, Emmanouil and Nam, Juhan and Mitsufuji, Yuki},
  title         = {Twenty-Five Years of {MIR} Research: Achievements, Practices, Evaluations, and Future Challenges},
  year          = {2025},
  eprint        = {2511.07205},
  archivePrefix = {arXiv},
  primaryClass  = {cs.SD}
}

@article{petersonberger1975,
  title={Cycles in symbol production: The case of popular music},
  author={Peterson, Richard A and Berger, David G},
  journal={American sociological review},
  pages={158--173},
  year={1975},
  publisher={JSTOR}
}

@article{prey2020curatorial,
  title={Locating power in platformization: Music streaming playlists and curatorial power},
  author={Prey, Robert},
  journal={Social media+ society},
  volume={6},
  number={2},
  pages={2056305120933291},
  year={2020},
  publisher={SAGE Publications Sage UK: London, England}
}

@article{prindle2003no,
  title={No competition: How radio consolidation has diminished diversity and sacrificed localism},
  author={Prindle, Gregory M},
  journal={Fordham Intell. Prop. Media \& Ent. LJ},
  volume={14},
  pages={279},
  year={2003},
  publisher={HeinOnline}
}

@book{rawls,
  title={A theory of justice. Rawls},
  author={Rawls, John},
  year={1971},
  publisher={The Belknap}
}

@article{serra2012,
  title={Measuring the evolution of contemporary western popular music},
  author={Serr{\`a}, Joan and Corral, {\'A}lvaro and Bogu{\~n}{\'a}, Mari{\'a}n and Haro, Mart{\'\i}n and Arcos, Josep Ll},
  journal={Scientific reports},
  volume={2},
  number={1},
  pages={521},
  year={2012},
  publisher={Nature Publishing Group}
}

@inproceedings{sturm2019,
  title={Artificial intelligence and music: open questions of copyright law and engineering praxis},
  author={Sturm, Bob LT and Iglesias, Maria and Ben-Tal, Oded and Miron, Marius and G{\'o}mez, Emilia},
  booktitle={Arts},
  volume={8},
  number={3},
  pages={115},
  year={2019},
  organization={MDPI}
}

@inproceedings{vickers2010loudness,
  title={The loudness war: Background, speculation, and recommendations},
  author={Vickers, Earl},
  booktitle={Audio Engineering Society Convention},
  volume={129},
  number={11},
  year={2010}
}

@article{hallam,
  title={The power of music: Its impact on the intellectual, social and personal development of children and young people},
  author={Hallam, Susan},
  journal={International journal of music education},
  volume={28},
  number={3},
  pages={269--289},
  year={2010},
  publisher={Sage Publications Sage UK: London, England}
}

@article{writers,
  title={Writers Guild of America Strike 2023},
  author={Dwivedi, Manish Kumar and Chhonkar, Aman},
  journal={Emerging Economies Cases Journal},
  pages={25166042251397697},
  year={2025},
  publisher={SAGE Publications Sage India: New Delhi, India}
}

@misc{ara2024,
  author       = {{Artist Rights Alliance}},
  title        = {Stop Devaluing Music: An Open Letter to the Music Industry},
  year         = {2024},
  howpublished = {\url{https://musically.com/2024/04/02/more-than-200-artists-sign-stop-devaluing-music-open-letter-on-ai/}},
  note         = {Accessed: 2026-04-01}
}

@misc{uscopyright2025,
  author       = {{US Copyright Office}},
  title        = {Copyright and Artificial Intelligence, Part 2: Copyrightability},
  year         = {2025},
  howpublished = {\url{https://www.copyright.gov/ai/Copyright-and-Artificial-Intelligence-Part-2-Copyrightability-Report.pdf}},
  note         = {Accessed: 2026-05-15}
}

\clearpage

\section{Supplementary Tables}
\begin{table}[!htbp]
\small
\centering
\caption{Summary of homogenization diagnostics and statistical signatures.}
\label{tab:diagnostics}
\begin{tabular}{@{}p{0.22\linewidth}p{0.25\linewidth}p{0.25\linewidth}p{0.18\linewidth}@{}}
\toprule
\textbf{Diagnostic} & \textbf{What it measures} & \textbf{Statistics reported} & \textbf{Signature of contraction} \\
\midrule
\textbf{D1}: Global dispersion
& Overall spread in 72-D acoustic space
& Track-to-centroid distances (Euclidean); Mann--Whitney U (one-sided); Cliff's $\delta$; label-shuffle permutation (10k); bootstrap CI (10k)
& Lower AI distances \\

\textbf{D2}: Feature variances
& Contraction of individual features
& Variance ratios (AI/Human); Levene test with BH-FDR ($q\leq 0.05$)
& Ratios $< 1$ \\

\textbf{D3}: Entropy
& Redundancy / predictability of feature use
& Shannon entropy ratios (AI/Human), histogram-binned ($k$=20)
& Ratios $< 1$ \\

\textbf{D4}: Geometric coverage
& Extent of occupied region under PCA projection
& PCA (up to 10 PCs); PC range ratios; convex-hull volume ratio (first $\leq$5 PCs); participation ratio ($D_\text{eff}$)
& Reduced AI coverage \\

\textbf{D5}: Separability
& Systematic acoustic differences
& 5-fold stratified CV accuracy/AUC (LR, RF, GBM, RBF-SVM); fold-wise scaling; feature importances (RF)
& High AI--human separation \\
\bottomrule
\end{tabular}
\end{table}

\begin{table}[!htbp]
\small
\centering
\caption{Within-genre mean pairwise distance, normalized to the human baseline (ratio $< 1$ = AI more homogeneous than humans within that genre). Lyria consistently falls near or below 1.0; Suno exceeds 1.0 in three of four genres.}
\label{tab:pairwise_distances}
\begin{tabular}{@{}lccc@{}}
\toprule
\textbf{Genre} & \textbf{Suno / Human} & \textbf{Lyria / Human} & \textbf{Human (raw)} \\
\midrule
Heavy Metal  & 1.004 & 0.773 & 9.43 \\
Afrobeats    & 1.582 & 0.832 & 8.81 \\
Dance Pop    & 1.272 & 1.119 & 7.64 \\
K-pop        & 1.382 & 0.972 & 7.99 \\
\bottomrule
\end{tabular}
\end{table}

\begin{table}[h]
\small
\centering
\caption{Genre separation ratio (between-genre centroid distance / within-genre spread). Higher values indicate more distinct genre categories. Suno collapses genre boundaries ($-36\%$ vs.\ human); Lyria preserves them.}
\label{tab:genre_separation}
\begin{tabular}{@{}lcc@{}}
\toprule
\textbf{System} & \textbf{Separation ratio} & \textbf{vs.\ Human} \\
\midrule
Human  & 0.662 & --- \\
Lyria  & 0.686 & $+1\%$ \\
Suno   & 0.442 & $-36\%$ \\
\bottomrule
\end{tabular}
\end{table}

\begin{table}[!htbp]
\small
\centering
\caption{Within-genre pairwise distance ratios (AI/Human) under MIR-steered (Exp.~1) and null-prompt (Exp.~2) conditions. Values below 1.0 indicate AI tracks are more homogeneous than the human baseline within that genre.}
\label{tab:null_comparison}
\begin{tabular}{@{}lcccccc@{}}
\toprule
& \multicolumn{3}{c}{\textbf{Suno}} & \multicolumn{3}{c}{\textbf{Lyria~3}} \\
\cmidrule(lr){2-4} \cmidrule(lr){5-7}
\textbf{Genre} & \textbf{MIR} & \textbf{Null} & \textbf{Diff} & \textbf{MIR} & \textbf{Null} & \textbf{Diff} \\
\midrule
Heavy Metal & 1.004 & 0.981 & $-0.023$ & 0.773 & 0.595 & $-0.178$ \\
Afrobeats   & 1.582 & 1.280 & $-0.302$ & 0.832 & 0.730 & $-0.102$ \\
Dance Pop   & 1.272 & 1.366 & $+0.094$ & 1.119 & 1.071 & $-0.048$ \\
K-pop       & 1.382 & 1.311 & $-0.071$ & 0.972 & 0.978 & $+0.006$ \\
\bottomrule
\end{tabular}
\end{table}

\begin{table}[!htbp]
\small
\centering
\caption{Title diversity across 100 Suno null-prompt tracks per genre. ``Unique titles'' counts distinct title strings; ``top title'' is the most frequent. Lyria~3 does not generate titles.}
\label{tab:titles}
\begin{tabular}{@{}lccp{3cm}@{}}
\toprule
\textbf{Genre} & \textbf{Unique titles} & \textbf{\% unique} & \textbf{Top title (count)} \\
\midrule
Afrobeats  & 13 & 13\% & ``Palm Wine Drift'' (20) \\
K-pop      & 16 & 16\% & ``Neon Orbit'' (10) \\
Dance Pop  & 15 & 15\% & ``Mirror Pulse'' / ``Anvil Breaker'' (14 each) \\
Heavy Metal & 14 & 14\% & ``Anvil Breaker'' (18) \\
\bottomrule
\end{tabular}
\end{table}

\FloatBarrier

\clearpage
\section{Supplementary Figures}

\begin{figure}[!htbp]
    \centering
    \includegraphics[width=0.85\linewidth]{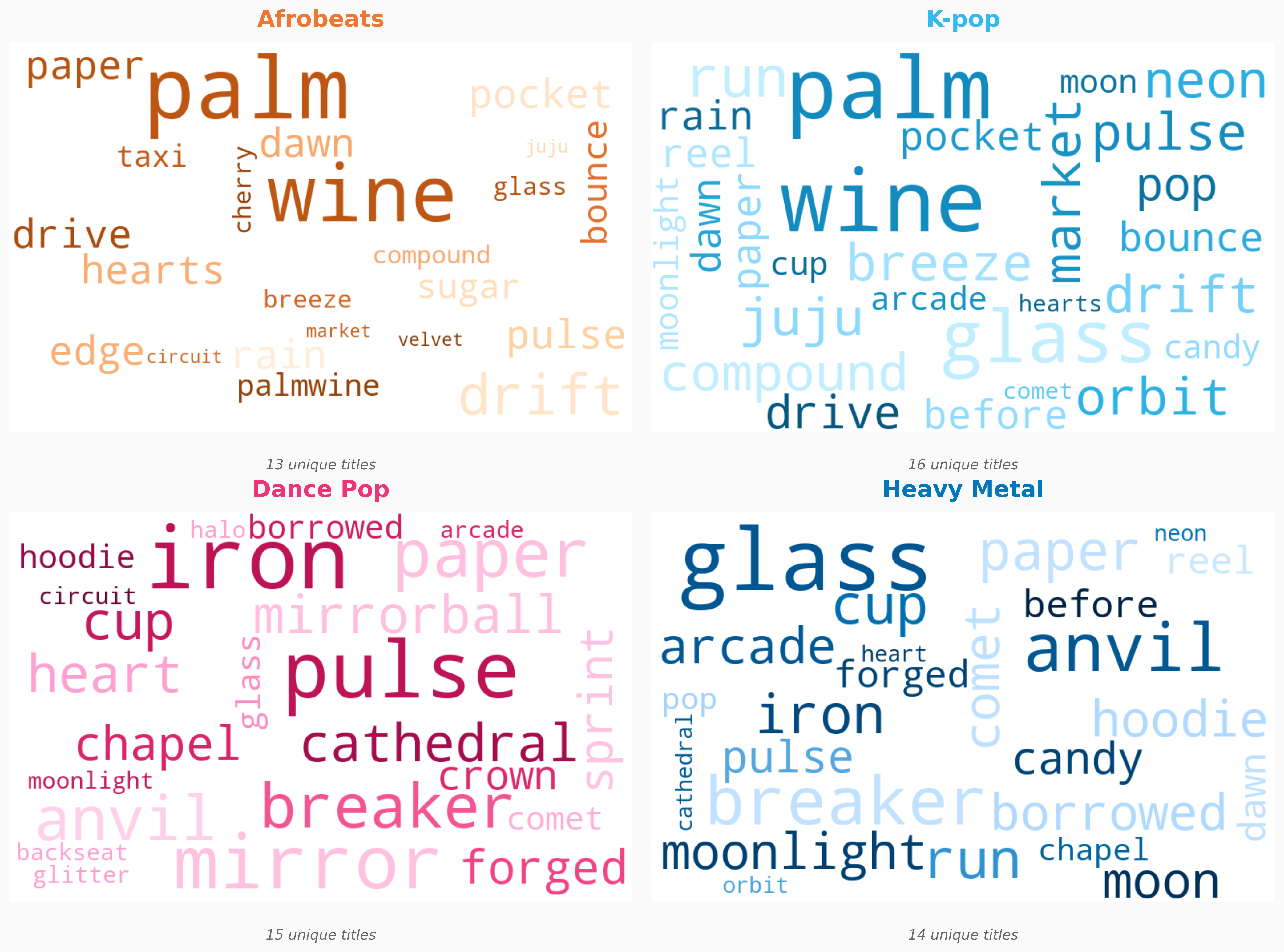}
    \caption{\textbf{Word cloud of Suno null-prompt titles (100 tracks per genre).} Word size proportional to frequency; color indicates the genre in which the word is most frequent (Afrobeats = amber, K-pop = pink, Dance Pop = blue, Heavy Metal = grey). The vocabulary is narrow and several words (e.g., ``Anvil,'' ``Pulse'') appear prominently across multiple genres, illustrating cross-genre leakage in title generation.}
    \label{fig:wordcloud}
\end{figure}

\end{document}